\documentclass[twocolumn,prd,aps,floats,nofootinbib,showpacs]{revtex4}

\usepackage{natbib}
\usepackage{graphicx}
\usepackage{epsfig}
\usepackage{rotate}
\usepackage{latexsym}
\usepackage{amssymb}
\usepackage{psfrag}
\usepackage{color}

\newcommand{\beq}{\begin{equation}}
\newcommand{\eeq}{\end{equation}}
\newcommand{\lsi}{\,\raisebox{-0.13cm}{$\stackrel{\textstyle<}
{\textstyle\sim}$}\,}

\def\<{\langle}
\def\>{\rangle}

\def\Ms{\,M_{\odot}}
\def\Ls{\,L_{\odot}}

\begin{document}

\title{\bf Implications of the Intermediate Mass Black Hole in globular cluster 
G1 on Dark Matter detection}
\author{Gabrijela Zaharija\v s}
\affiliation{{\it HEP Division, Argonne National Laboratory, 9700 Cass Ave., 
Argonne, IL 60439}}
\begin{abstract}
Recently there has been a growing evidence in favor of the presence of an 
Intermediate Mass Black Hole in the globular cluster G1, in Andromeda Galaxy. 
Under the assumption that formation of this globular cluster occurred within a 
dark matter halo, we explore whether the presence of a black hole could result in an observable 
gamma ray signal due to dark matter annihilation in this globular 
cluster. Starting from an initial NFW matter profile, with density parameters 
consistent with G1 observations, we find that indeed, if the spike in the 
density has been formed and has survived till present, the signal could be 
observed by GLAST and current ACT detectors. 
\end{abstract}

\pacs{95.35.+d \hfill ANL-HEP-PR-08-15}

\maketitle

%\vspace{4pt}

\section{Introduction}
One of the promising ways of detecting self-annihilating dark matter is through 
its final 
annihilation products, gamma rays being one of them. These, so called 
indirect detection techniques, offer an opportunity to identify dark matter (DM) 
properties, complementary to other searches, as dark matter 
direct detection and accelerator experiments. 
Gamma rays with energies higher than $1$ GeV have been searched for, in both 
satellite 
based (EGRET \cite{egret}) and ground-based experiments (HESS \cite{hess}, MAGIC 
\cite{magic}, CANGAROO \cite{cangaroo}, VERITAS \cite{veritas},...) and new 
telescopes are being planned and built, such as GLAST \cite{glast} with launch scheduled 
for the next spring. 
The gamma ray signal of self-annihilating dark matter depends on the 
density squared and therefore ``hot spots'' where DM density is enhanced, are 
expected to exist across the sky. 
Promising sites for gamma ray DM detection should, therefore, have high DM density, be 
located in our vicinity and have a low 
level of astrophysical background, as the centers of our own and nearby 
galaxies, dwarf galaxies of the Milky 
Way, earth sized halos or primordial intermediate-mass black holes, for a review see \cite{Bertone:2004pz}. 

In this paper we propose yet another site for dark matter indirect 
observation, the globular cluster G1 in the Andromeda Galaxy. This cluster has 
recently been observed with X-ray \cite{Kong:2007mu} and radio \cite{Ulvestad:2007cw} telescopes and both 
measurements detected a source in the cluster core, consistent with the 
presence of an Intermediate Mass Black Hole (IMBH). Dynamical considerations, 
based on the most recent HST/WFPC2 and KECK/HIRES spectroscopic measurements 
support an 
IMBH presence, \cite{Gebhardt:2005cy} (see also, 
\cite{Gebhardt:2002in,Baumgardt:2002rc,Baumgardt:2003an}). In this letter we 
explore whether the presence of an IMBH could have an effect on the DM density 
favorable enough for G1 to be a promising target for indirect detection. 
Dark matter halos react to the growth of black holes. In the case of 
adiabatic growth of a central black hole, the density is strongly enhanced in 
its vicinity, resulting in a spike \cite{Gondolo:1999ef}. Since the gamma ray 
flux depends on the square of the density, a spike would enhance the flux by 
several orders of magnitude. The existence and the slope of a spike is 
highly uncertain, however. We will later see how spikes with quite 
shallow slopes could still produce an observable signal.    

G1 as a dark matter observation target suffers from being rather far from the 
Earth, but as we will see below, the enhancement of the density due to a 
black hole can potentially compensate for the 
distance. For example, if compared to the dwarf galaxy Draco, G1 is 10 times 
farther, but the signal from G1, in the case of a spike could be $10^{2\div 
4}$ times stronger than the one expected from Draco, \cite{Bergstrom:2005qk}. 
Also, on the positive side, globular clusters are not expected to 
have significant background in high-energy gamma rays, as we will argue later. 
Relic IMBH have long been considered good sites for DM detection, but suffered 
from 
uncertainty in their location. If G1 
really hosts an IMBH it would offer a unique opportunity to study these 
objects and their interaction with dark matter.

\section{Globular clusters and G1}

{\it Globular Clusters} are very dense stellar systems: they have the typical 
mass of dwarf galaxies, but the size a factor of $\sim 10$ smaller. They are 
found in the halo or bulge regions of galaxies, \cite{1543}. Because they are 
old, they do not have significant amounts of interstellar 
gas and the stars within globular cluster are usually coeval. This suggests 
that the astrophysical background in high energy gamma ray telescopes should be 
small. 

In the primordial scenario of globular cluster formation \cite{Peebles:1968nf},
globular clusters formed in DM minihaloes, before or during reionization, being 
the first galaxies to form; It is plausible, therefore, that globulars 
initially had a dominant dark matter component. They were then accreted by larger galaxies and lost 
their extra nuclear material due to stripping by the tidal forces.  Today globular clusters 
are baryon dominated, with $M/L$ of a few. They have very dense stellar cores 
and the presence of 
tidal tails observed on globular cluster Palomar 5 suggests that they do not
 have an extended DM halo \cite{Moore:1995pb}. How much dark matter could be 
present in globular clusters? King-Michie 
models \cite{1979AJ.....84..752G} do a satisfactory job in the dynamical 
description of 
globular clusters while considering only its stellar population. In that sense, 
presence 
of exotic dark matter is not required and no lower 
bound on DM annihilation signal from G1 can be placed. However, the simulations 
of 
Mashchenko and Sills \cite{Mashchenko:2004hk} of the evolution of globular 
clusters in tidal fields show that the 
dark matter present in their nucleus could be compressed and its density 
profile steepened by these dynamical processes and that, depending on the 
initial dark matter 
profile, and type of DM-stellar interaction, some globular clusters could still 
present an extended halo.

{\it The G1 cluster} is one of the most luminous and the most massive globulars 
in the Local Group. 
Its total luminosity  is $L\sim 2~10^6\Ls$ and its mass is estimated to be 
between $1.4$ and $1.7 \times 10^7\Ms$ in King-Michie models, 
\cite{Meylan:2001yy}. However, the values of the total mass of the cluster has 
large 
scatter depending on the model used for its calculation. For example 
Baumgardt {\it et al.} \cite{Baumgardt:2003an} find a value of 
$(8\pm 1)~10^6\Ms$ in their simulation which uses an evolutionary model. 
The main reason for these discrepancies is that spectroscopic data on G1 are 
rather poor: measurements of the velocity dispersion profile in the cluster 
was performed only in the inner region of the cluster (within $\sim 5$ pc, 
\cite{Gebhardt:2005cy}) or the velocity dispersion was reduced to a single 
value, measured within a slit $\sim 6\times 24$ pc, \cite{1997ApJ...474L..19D}.

G1 has a stellar core radius of $0.52$ pc, and a half mass 
radius of $13$ pc; its tidal radius is about $200$ pc, \cite{Meylan:2001yy}. The 
cluster lies at a distance of $770$ kpc from us and it would be
clearly separated from the bright Andromeda's disk in current high energy gamma ray telescopes.

G1 has several properties not typical for globular clusters. It has a very 
large central velocity dispersion \cite{1997ApJ...474L..19D}, has a very 
flattened shape, with 
mean ellipticity of $0.2$ and it shows a spread in metal abundances among its 
stellar population. In fact, one of the possible scenarios of G1 formation is 
that it is a surviving nucleus 
of a dwarf elliptical galaxy, which would have lost 
its outer envelop through tidal interaction with M31. In \cite{Bekki:2004tn}, Bekki and Chiba show that also in this scenario presence of DM in the 
core is plausible.

{\it Evidence for a black hole in G1: }%\label{sec:Evidence} 
While there is a growing knowledge of super-massive and 
stellar mass black holes, intermediate mass black holes are yet poorly 
understood. Recent detections of 
AGNs in low-luminosity, late-type galaxies suggest that IMBHs do exist (see, 
for instance, \cite{Filippenko:2003kg}). 

Several papers dealt with a dynamical analysis of the G1 cluster and the 
possible presence of an IMBH with different outcomes \cite{Gebhardt:2005cy,Baumgardt:2003an,Baumgardt:2002rc}. 
The most recent analysis by Gebhardt {\it et al.} \cite{Gebhardt:2005cy} (G05), favors its presence. They find the best 
fit black hole mass to be $1.8\pm 0.5 \times 10^4\Ms$.

X-ray observations with XMM Newton reported a source with 
luminosity expected from an accretion onto an IMBH \cite{Pooley:2006aq}. Radio 
observations, using Very Large Array detected a faint radio source 
within an arcsecond of the cluster core \cite{Ulvestad:2007cw} with luminosity consistent with mass estimates 
by G05 and measured X-ray luminosity, thereby strengthening the evidence for a 
black hole presence.

\begin{figure}[t]
\begin{center}
\includegraphics[height=5.cm,width=.65\columnwidth]{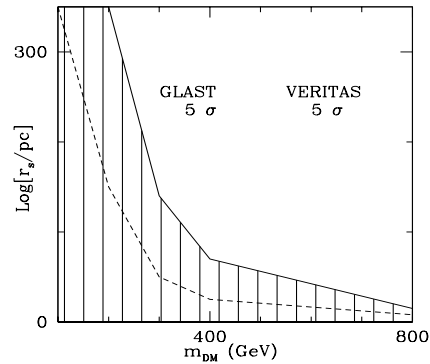}
\caption{\label{figure} The non-shaded region is the range of $r_s$ and $m_{DM}$ 
which could be detectable by GLAST or VERITAS with $5\sigma$ significance or 
better. Shaded region is the excluded region from the EGRET observation. Solid 
line represents the case of $\Delta M=9~10^4 \Ms$, and dotted one $\Delta 
M=3~10^4 \Ms$.}
\end{center}
\end{figure}

\section{Dark Matter density profile}
\label{sec:DMann}

We consider a cuspy NFW dark matter profile, $\rho (r)=\rho _0 (r/r_s)^{-1}
(1+r/r_s)^{-2}$
where $\rho_0$ is central density and $r_s$ the scale radius.

Dark matter halos react to the growth of black holes, and in the case of 
adiabatic growth, the result is the formation of large over-densities called 
spikes, \cite{Gondolo:1999ef}. If the initial dark matter density was cuspy, 
after the adiabatic growth, the 
dark matter assumes a distribution given by
\beq
\label{gammasp}
\rho _{sp,NFW} (r)=\rho _{cusp}(R_{sp})~g_{\gamma} (r) 
\left( \frac{R_{sp}}{r} \right)^{\gamma _{sp}},
\eeq
where $R_{sp}$ is the radius of influence of the black hole, found as a 
solution of the equation $\int _{0} ^{R_{sp}} \rho(r)d^3r=2M_{BH}$
$g_{\gamma}(r)\sim (1-4R_{S}/r)^3$, where $R_{S}$ is the Schwarzchild 
radius of the black hole and $\gamma _{sp}=7/3$ in the NFW case.

On the other hand, in models with finite cores, the 
slope of the spike is shallower, $\gamma _{sp}=3/2$.  Numerical simulations of 
halo formation predict cuspy profiles, \cite{Bertone:2004pz}. However, simulations do not take into 
account the baryon-DM interactions and have other limitations, and we therefore consider a Burkert profile 
\cite{Burkert:1995yz}, which has a core at small $r$, $\rho (r)=\rho _0(1+r/r_s)
^{-1}[1+(r/r_s)^2]^{-1}$. We parametrize the final dark matter profile after 
the growth of the spike 
as $\rho _{sp,core} (r)=\rho _{0,core}~g_{\gamma} (r) (R_{sp}/r)^{3/2}.$ 

Due to annihilations in the inner regions of the spike, there will exist a 
maximal dark matter density, given by $\rho _{max}=m/\< \sigma v \> t_{BH}$, 
where $m$ is dark matter mass, $\<\sigma v\>$ is the thermally averaged 
annihilation cross section and $t_{BH}$ is the age of IMBH, which we take to be 
$10^{10}$ years.
The final dark matter profile is then given by $\rho _{sp,f} (r)=\rho_{sp}~\rho 
_{max}/(\rho _{sp}+\rho _{max})$.

The adiabatic growth approach used here maximizes the effects of the black hole 
on the dark matter density profile. Substantial work has been done on a more 
realistic description of black hole influence on dark matter evolution, and the 
existence of a spike in our Galactic Center has been questioned, \cite{imbh}. Various processes, such as off-center black hole 
formation, black hole merging events, dark matter scattering by stars, etc. 
could all result in slopes shallower than predicted by
the adiabatic growth model. The reconstruction of the central mass profile of G1, in G05, points to the 
existence of a cusp, both in the central luminosity and in the mass 
profiles. This in turn suggests that G1 did not go through a recent merger, 
which would deplete both stellar and dark matter cusps. On the other hand, 
stellar-dark matter scatterings might be important in G1.  Gravitational scattering of dark matter with stars could deplete 
the spike and the slopes are generically well described with slope $3/2$, if this effects are taken into account,  \cite
{Merritt:2003qk}. To probe this effect 
we somewhat artificially vary $\gamma_{sp}$ and find the lowest value which 
would result in a detectable signal. 

To construct the DM density in G1 we start from the assumption that the outer 
part of the dark matter halo has been stripped together with the stellar 
envelope, while the dark matter in the center has survived due to the 
compactness of the core, as motivated by above mentioned simulations, \cite
{Bekki:2004tn,Mashchenko:2004hk}. We conservatively truncate the profile at the 
tidal radius. Since the main contribution to the annihilation signal comes from the inner 
region of the profile ($r\lsi r_s$) this assumption would not significantly 
affect the result. Also, we conservatively assume that the inner slope of the 
density profile does not change with the evolution of the cluster.

The two parameters of the density distribution, $r_s$ and $\rho_0$, are poorly 
constrained in globular clusters. One typically constrains $r_s$ and $\rho_0$ by 
solving spherical Jeans equation, based on the  velocity dispersion profile 
measurement, \cite{Mashchenko:2005bj,Strigari:2006rd}. In the case of 
G1, G05 find the mass density profile in the inner 5pc, using this approach
(the  ellipticity of G1 is less than $0.1$ in the innermost region of the cluster, \cite{Meylan:2001yy}, 
suggesting that the spherical approximation might be valid there), but 
determination of $r_s$ and $\rho _0$ is further 
complicated by the fact that G1 is baryon dominated, and that the stellar 
$M/L$ dependence on the radius is not well understood. We choose $r_s$ and 
$\rho _0$ by making sure that the dark matter density profile satisfies the 
three constraints listed below:
 
i) We use the uncertainty on the total cluster mass $\Delta M$ as the limit on 
the 
amount of dark matter allowed in the cluster:
$\int \rho _{DM} d^3 {\vec r}=\Delta M$. As $\Delta M$ we use both the error on 
the total mass within the King-Michie model ($3~10^6 M_{\odot}$) or the mass 
dispersion between various models ($9~10^6 M_{\odot}$).

ii) When adiabatic compression due to the black hole is present, the strongest 
limit might come from the measurement uncertainty of the total mass measured in 
the first bin ($\sim 0.1$ pc), which is $\sim 3~10^3$ $M_{\odot}$, \cite
{Gebhardt:2005cy}. We use this mass as an upper limit on the amount of DM in the 
central $\sim 0.1$ pc; 
 
iii) We use the mass density profile, as found by G05, by imposing that the dark matter density 
should not be bigger than $10\%$ of the found mass density of the cluster at any 
distance (10$\%$ is somewhat arbitrary value,  error bars on the mass 
density of the cluster as calculated in G05, are hard to estimate; the 
measurement error in the central bin is $15\%$). 

For comparison, we notice, that the values of $r_s$ (and $\rho_0$) can be 
estimated also by using the virial mass $M_{vir}$ and the concentration 
parameter $c_{vir}$, (for details, see \cite{Ullio:2002pj}). If we take for 
$M_{vir}$ to be $\sim 5$ times the mass of G1 today, it leads to a value of $r_s$ for the NFW profile of $\sim 270$ pc, which is close 
to its tidal radius today.

\section{Detection signal}
\label{sec:Signal}

In this section we estimate the strength of the dark matter annihilation 
signal as expected in the GLAST and ground-based 
Atmospheric Cerenkov Telescopes (ACTs). The EGRET experiment did not measure 
any signal above the diffuse gamma ray background in the direction of G1. 
We use this fact to place further limits on the value of $r_s$ and $\rho_0$. 

In the case of G1, when the typical scale of the dark matter halo is much smaller than 
the distance to the object, the flux of gamma rays produced by 
self-annihilating dark matter is given by, \cite{Bergstrom:2005qk},
\beq
\label{flux}
\Phi =  \frac {N_{\gamma}}{4\pi D^2}\int dr 4 \pi r^2 
\frac{\langle \sigma v\rangle }{2}\left( \frac{\rho (r)}{m_{dm}} \right)^2
\eeq
where $D$ is distance to the object and $m_{dm}$ is the dark matter 
mass;  $N_{\gamma}$ is number of photons produced per annihilation and it is 
calculated here by means of the analytic fitting formula for the $\gamma$ ray 
spectrum produced by particles which annihilate dominantly to gauge bosons \cite
{Bergstrom:1997fj} which is a sufficiently accurate approximation for our 
purposes; we assumed a value for DM annihilation cross section $\langle \sigma v 
\rangle$ of  $3~10^{-26}$ cm$^3$s$^{-1}$, corresponding to the value expected 
for thermally decoupled relic.    

For GLAST experiment we assume exposure of 1 m$^2$yr. The biggest contribution 
to the background for G1 comes 
from the Galactic diffuse emission at intermediate latitudes. Its strength is $\lsi 6~10^{-6} ({\rm GeV}/{\rm E})^{-2}
$ cm$^{-2}$ s$^{-1}$ sr$^{-1}$ GeV$^{-1}$, \cite{Strong:2004de}. Using field of view of $10^{-5}$ sr, this 
background accumulates to about $20$ events in the given exposure.

For ACT experiments we assume an effective area of $4~10^8$ cm$^2$, 40 hours 
of observation and an energy threshold of 100 GeV. The dominant background in 
ACTs comes from hadronic showers which are misidentified as gamma rays. Its 
spectrum is given as ${dN_{had}}/{dE}\sim 3~\epsilon _{had} 
({E}/{\rm GeV} ^{-2.7})$ GeV$^{-1}$ sr$^{-1}$ cm$^{-2}$ s$^{-1}$, \cite
{Bergstrom:2005qk}, where $\epsilon _{had}$ is the fraction of misidentified 
hadronic showers. For $\epsilon _{had}=0.01$, typical for ACTs, and a field of 
view of $10^{-5}$ sr, this background accumulates at a rate of $\sim 80$ events 
per hour of exposure.

In the direction of G1, EGRET observed flux of $\sim 5~10^{-7}$ cm$^2$s$^{-1}$sr
$^{-1}$ above 1 GeV, \cite{egret}. The area of EGRET was $6400$ cm$^2$, the 
angular 
resolution $0.5^\circ$ and we will assume one year of effective time. 
The measured number of events, from the direction of G1, in the energy range 
1-10 GeV is $\sim 25$.

We show the results in a Figure \ref{figure}, where a NFW profile is assumed. 
The lower shaded region is excluded by EGRET. The upper right corner of the 
figure is observable by either GLAST or VERITAS experiments, with $5 \sigma$ 
significance, or better. We did not consider values of $r_s$ higher than 300 pc, since the tidal radius of G1 is 
200 pc. We comment here 
that spikes with slopes as shallow as $\gamma _{sp}=1.75$ could still be 
observable, albeit for a very small parameter space: say for $m=100$ GeV, the 
observable range of $r_s$ would be 3 pc $\lsi r_s \lsi 5.2$ pc  (for $\Delta 
M=3~10^6 \Ms$), but with the parameter space enlarging considerably for higher $
\gamma _{sp}$. This is meant to illustrate that even slopes shallower than the 
ones predicted by adiabatic growth of a black hole, could potentially be 
observable. We did not perform a detailed analysis in this case, instead we 
varied ad hoc the values of $\gamma_{sp}$ in Eq. (\ref{gammasp}) in order to 
mimic complicated superposition of effects in the central DM region (initial 
steepening of DM cusp, motivated by \cite{Mashchenko:2004hk}, and various 
dynamical processes that might contribute to the depletion of the spike). If the 
dark matter profile contains no spike, we find that its selfannihilation signal 
would be unobservable by GLAST and current ACTs. We also find that the core 
profile would not produce an observable signal even if the adiabatic spike is 
present. The main reason for this is the combined effect of a shallow slope of a 
spike and strong constraints on DM density profile parameters in the case of 
core profile.

\section{Summary}
\label{sec:Summary}

In this paper we show that the dark matter annihilation signal from the globular
cluster G1 in Andromeda could be within the range of GLAST and ACT experiments 
if the DM cusp has steepened due to the influence of an IMBH. 

G1 suffers from a high uncertainty in the determination of its dark matter 
density profile; more measurements of velocity dispersion profile are needed, 
as well as a better understanding of the stellar mass-to-light ratio. G1 is 
also an extragalactic source, a factor of $\sim 10$ farther then typical dwarf galaxies of the Milky Way. On the 
positive sides, it has potentially an IMBH in 
its center which could substantially enhance the dark matter signal. Because 
of its old stellar population, G1 should not have a high astrophysical 
background in high energy gamma rays.

With the GLAST detector coming up in the next year many new gamma ray sources 
will 
be discovered. It is possible that some of them will be uniquely 
interpreted as a dark matter signal. If we are lucky, the signal from G1 could add some information
in this respect and maybe help disentangling the dark matter-astrophysical 
properties.        

The author would like to kindly thank Dan Hooper and Robyn Levine with whom the 
idea for the paper emerged and to Pasquale Serpico and Emiliano Sefusatti for 
many helpful comments. Work supported in part by the US Department of Energy, 
Division of High Energy Physics, under
Contract DE-AC02-06CH11357.

%\bibliography{globwriteupSAVE}
%\bibliographystyle{unsrt}

\end{document}